\documentclass[a4paper,11pt]{article}
\pdfoutput=1 % if your are submitting a pdflatex (i.e. if you have
\usepackage{jheppub} % for details on the use of the package, please
\usepackage[T1]{fontenc} % if needed
\graphicspath{{fig/}}

\title{\boldmath Inclusive $J/\psi$ photoproduction at the ILC within the framework of non-relativistic QCD}

\author[a,b,1]{Xi-Jie Zhan,\note{Corresponding author.}}
\author[a,b]{Xing-Gang Wu,}
\author[a,b]{Xu-Chang Zheng}

\affiliation[a]{Department of Physics, Chongqing University,\\Chongqing 401331, People's Republic of China}
\affiliation[b]{Chongqing Key Laboratory for Strongly Coupled Physics, Chongqing University,\\Chongqing 401331, People's Republic of China}

\emailAdd{zhanxj@cqu.edu.cn}
\emailAdd{wuxg@cqu.edu.cn}
\emailAdd{zhengxc@cqu.edu.cn}

\abstract{Based on the non-relativistic quantum chromodynamics factorization framework, we study the inclusive $J/\psi$ photoproduction at the future high energy $e^+e^-$ collider, International Linear Collider(ILC), where the initial photons come from the back-scattering of laser and electron (positron). The intermediate states, $c\bar{c}$$({}^3\!S_1^{[\textbf{1}]}$,${}^1\!S_0^{[\textbf{8}]}$,${}^3\!S_1^{[\textbf{8}]}$,${}^3\!P_{0,1,2}^{[\textbf{8}]})$, and resolved photoproduction processes are considered. Numerical results show that the cross section of $J/\psi$ photoproduction at the ILC could be very large, and the single resolved process via $c\bar{c}({}^1\!S_0^{[\textbf{8}]})$ intermediate state dominates primarily the production. We also present various kinematic distributions for $J/\psi$ photoproduction. Combining the high luminosity of the collision, $J/\psi$ photoproduction at the ILC shall provide a good platform to test the NRQCD factorization. }

\begin{document}
\maketitle
\flushbottom

\section{Introduction}
\label{sec:1}

Heavy quarkonium consist of two quarks with heavy mass, since its discovery, it has been well probed to investigate the strong interaction and test quantum chromodynamics (QCD) due to its characteristic scales in the system. Non-relativistic QCD(NRQCD) factorization framework~\cite{Bodwin:1994jh} was proposed to describe the production and decay of heavy quarkonium. This effective theory introduces the color-octet (CO) mechanism and surpasses the traditional color-singlet (CS) model~\cite{Berger:1980ni, Baier:1981uk, Humpert:1986cy, Gastmans:1986qv, Gastmans:1987be}. NRQCD factorization has achieved great successes in the production~\cite{Campbell:2007ws, Gong:2008ft, Butenschoen:2010rq, Ma:2010yw} and polarization~\cite{Gong:2010bk, Butenschoen:2012px, Chao:2012iv, Gong:2012ug, Feng:2018ukp} of heavy quarkonium at the hadron colliders. The high-energy and high-luminosity electron-positron collider is considered as one of the main next generation colliders, such as the CEPC~\cite{CEPCStudyGroup:2018rmc, CEPCStudyGroup:2018ghi}, FCC-ee~\cite{FCC:2018evy}, ILC~\cite{ILC:2007bjz, Erler:2000jg} and so on. According to their designs, these colliders can run at various collision energies with high luminosity. Therefore it could be expected that future high-energy $e^+e^-$ colliders will provide excellent experimental platforms for the study of heavy quarkonium. In contrast to hadron colliders, the $e^+e^-$ collider has less background for the production of heavy quarkonium and theoretical calculation is simpler. At the $e^+e^-$ collider, the best way to produce heavy quarkonium via the electron-positron annihilation, especially when the collision energy is around $Z$-boson mass; and due to the resonance effect, high yield of heavy quarkonium is expected~\cite{Sun:2013liv}.

\begin{figure}[t]
	\centering % \begin{center}/\end{center} takes some additional
	\includegraphics[width=.6\textwidth]{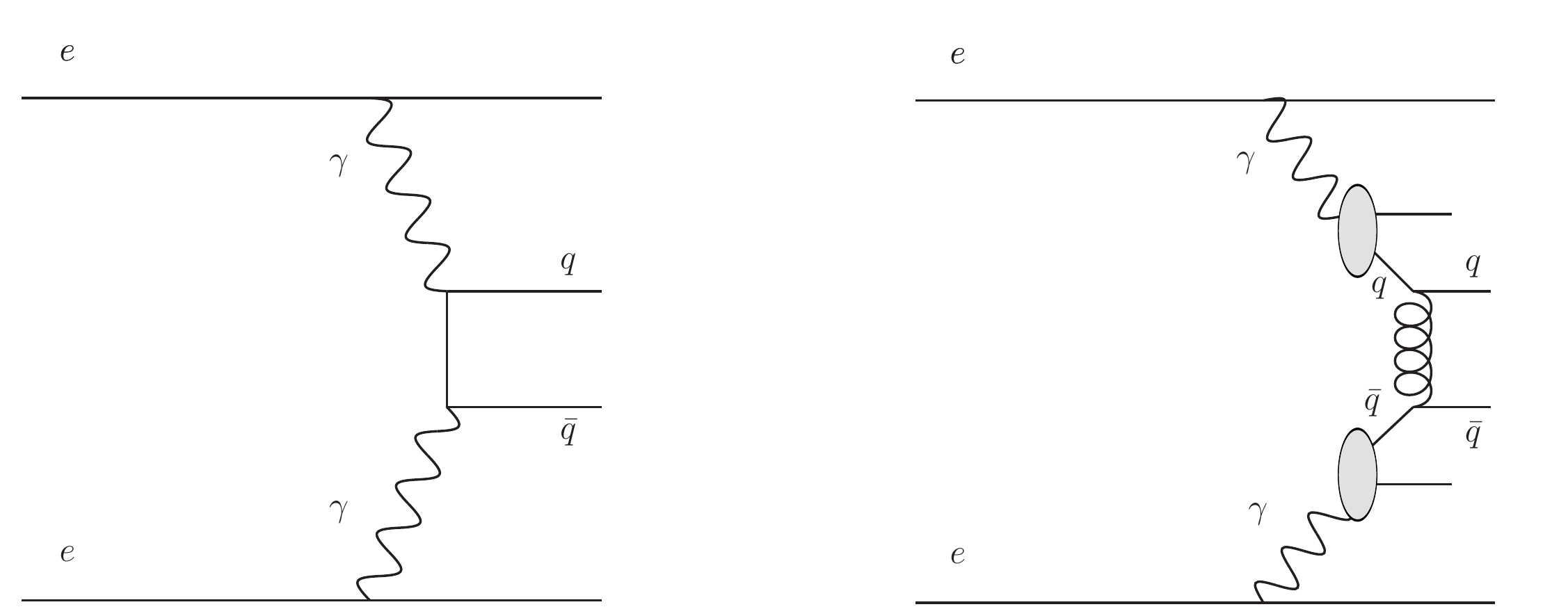}
	\caption{\label{fig:11} Diagrams for example of direct(left) and resolved(right) photoproduction processes at $e^+e^-$ collider. The diagrams are drawn by JaxoDraw~\cite{Binosi:2003yf}.}
\end{figure}

In addition to the annihilation mode, photoproduction at the $e^+e^-$ collider is also an important production channel of heavy quarkonium. There are two main sources of photons. For example, at the CEPC, photons can come from bremsstrahlung of initial electron and positron, which is described by Weiz\"acker-Williams approximation (WWA)~\cite{Frixione:1993yw}. At the ILC, photons can be scattered out by external laser and the electron beam. In the eyes of QCD, the photon could be in a hadronic state besides the bare photon state~\cite{Chu:2017mnm}. Due to quantum fluctuation, the photon can undergo a transition for a short period of time into a light quark pair or gluon. As a result, the photon as a whole particle, can participate directly the hard interaction to produce heavy quarkonium, which is called as the direct photoproduction. Alternatively, the photon resolves and the light quarks or gluons in the photon also get into the interaction. This is called as the resolved photoproduction, as shown in figure~\ref{fig:11}.

The $J/\psi$ photoproduction has been studied in many literature~\cite{Ma:1997bi, Japaridze:1998ss, Godbole:2001pj, Qiao:2001wv, Kniehl:2002wd, Klasen:2003zn, Artoisenet:2007qm, Klasen:2001mi, Klasen:2008mh, Li:2009zzu, Chen:2014xka, Sun:2015hhv, Chen:2016hju, Yang:2020xkl, Yang:2022yxb}. The inclusive $J/\psi$ photoproduction was measured in 2001 by the DELPHI Collaboration at CERN LEP II~\cite{TodorovaNova:2001pt,Abdallah:2003du}. Theoretical calculation of the $p_t$ distribution based on the CS model is smaller by one order of magnitude than the experimental measurements. After considering the CO mechanism, NRQCD gave nice description of the measurements~\cite{Klasen:2001cu} and this has been viewed as one of the earliest evidences of the existence of color-octet processes in nature. With the CO LDMEs extracted by a global fit of worldwide data~\cite{Butenschoen:2011yh}, however, the next-to-leading order NRQCD prediction of $J/\psi$ photoproduction systematically underestimate the DELPHI data. It is noteworthy that only a few $J/\psi$ events were reconstructed at LEP II and the measurements have large uncertainties~\cite{He:2019tig1}. At present there are no other experiments yet to verify the results of LEP II. As for the hadron collider, NRQCD factorization has achieved great success in explaining the experimental measurements of heavy quarkonium, but it still does not give a unified description of those observables, such as the total yield, kinematic distribution, and polarization~\cite{Chen:2021tmf}. Consequently it is necessary to measure the production of heavy quarkonium at other platforms such as the high-energy $e^+e^-$ collider so as to further test the NRQCD factorization framework.

In this work, we study the inclusive $J/\psi$ photoproduction at the ILC, including both the color-octet channels and the resolved photoproduction processes. In section~\ref{sec:2}, we give the basic theoretical framework of the calculation. The numerical results and discussions are presented in section~\ref{sec:3} and a brief summary is in section~\ref{sec:4}.

\section{Formulation and calculation}
\label{sec:2}

At the International Linear Collider, initial photons can achieve high energy and high luminosity and its spectrum is described as~\cite{Ginzburg:1981vm},
\begin{equation}
f_{\gamma/e}(x)=\frac{1}{N}\left[1-x+\frac{1}{1-x}-4 r(1-r)\right],
\end{equation}
where $x=E_{\gamma} / E_{e}$, $r=x /\left[x_{m}(1-x)\right]$, and the normalization factor,
\begin{equation}
N=\left(1-\frac{4}{x_{m}}-\frac{8}{x_{m}^{2}}\right) \log(1+x_m)+\frac{1}{2}+\frac{8}{x_{m}}-\frac{1}{2 (1+x_m)^{2}}.
\end{equation}
Here $x_{m}=4 E_{e} E_{l} \cos ^{2} \frac{\theta}{2}$, $E_e$ and $E_l$ are the energies of incident electron and laser beams, respectively, and $\theta$ is the angle between them. The energy of the  laser backscattering (LBS) photon is restricted by
\begin{equation}
0 \leq x \leq \frac{x_{m}}{1+x_{m}},
\end{equation}
with optimal value of $x_m$ being $4.83$~\cite{Telnov:1989sd}.

Within the NRQCD factorization framework, the production cross section of heavy quarkonium is separated into a product of the long distance matrix elements (LDMEs) and the short distance coefficients (SDCs). The SDC describes the production of an intermediate $Q\bar{Q}(n)$-pair, where the quantum number $n={}^{2S+1}\!L_J^{[c]}$ with $c$ being color multiplicity. The LDME describes the hadronization of the $Q\bar{Q}(n)$-pair into heavy quarkonium. The SDCs can be calculated perturbatively while the LDMEs are assumed to be process-independent and can be extracted by global fitting of experimental data. The differential cross section for  heavy quarkonium ($H$) photoproduction is then formulated as double convolution of the cross section of parton-parton (or photon) processes and the corresponding parton distribution functions,
\begin{equation}
\begin{aligned}
\mathrm{d} \sigma &  \left(e^{+} e^{-} \rightarrow e^{+} e^{-} H+X\right) \\
= &\int \mathrm{d} x_{1} f_{\gamma / e}\left(x_{1}\right) \int \mathrm{d} x_{2} f_{\gamma / e}\left(x_{2}\right) \sum_{i, j, k} \int \mathrm{d} x_{i} f_{i / \gamma}\left(x_{i}\right) \int \mathrm{d} x_{j} f_{j / \gamma}\left(x_{j}\right) \\&
 \times  \sum_{n} \mathrm{~d} \sigma(i j \rightarrow c \bar{c}[n]+k)\left\langle \mathcal{O}^{H}[n]\right\rangle,
\end{aligned}
\end{equation}
here, $f_{i/\gamma}$ is the Gl\"uck-Reya-Schienbein(GRS) parton distribution functions of the light quarks and gluon in photon~\cite{Gluck:1999ub}.
$d\sigma(ij\to c\overline{c}[n]+k)$ represents the differential partonic cross section, $i,j=\gamma,g,q,\bar{q}$ and $k=g,q,\bar{q}$ with $q=u,d,s$.
$c\overline{c}[n]$ are the intermediate $c\overline{c}$ pair with states $n={}^3\!S_1^{[\textbf{1}]},{}^1\!S_0^{[\textbf{8}]},{}^3\!S_1^{[\textbf{8}]},{}^3\!P_J^{[\textbf{8}]}$ for $H=J/\psi,\psi(2S)$ and $n={}^3\!P_J^{[\textbf{1}]},{}^3\!S_1^{[\textbf{8}]}$ for $H=\chi_{cJ}$($J=0,1,2$), respectively. $\langle{\cal O}^H[n]\rangle$ are the LDMEs of $H$.

Heavier charmonia, such as $\psi(2S)$ and $\chi_{cJ}(J=0,1,2)$, can decay into $J/\psi$. These feed-down contributions are taken into account by multiplying their direct-production cross sections with corresponding decay branching ratios to $J/\psi$,
\begin{equation}
\begin{aligned}
d \sigma^{\text {prompt} J / \psi} = & d \sigma^{J / \psi}+ d \sigma^{\psi(2 S)} B r(\psi(2 S) \rightarrow J / \psi+X)\\
&+\sum_{J} d \sigma^{\chi_{c J}} B r\left(\chi_{c J} \rightarrow J/\psi+\gamma\right).
\end{aligned}
\end{equation}

\begin{figure}[t]
	\centering
	\includegraphics[width=0.24\textwidth]{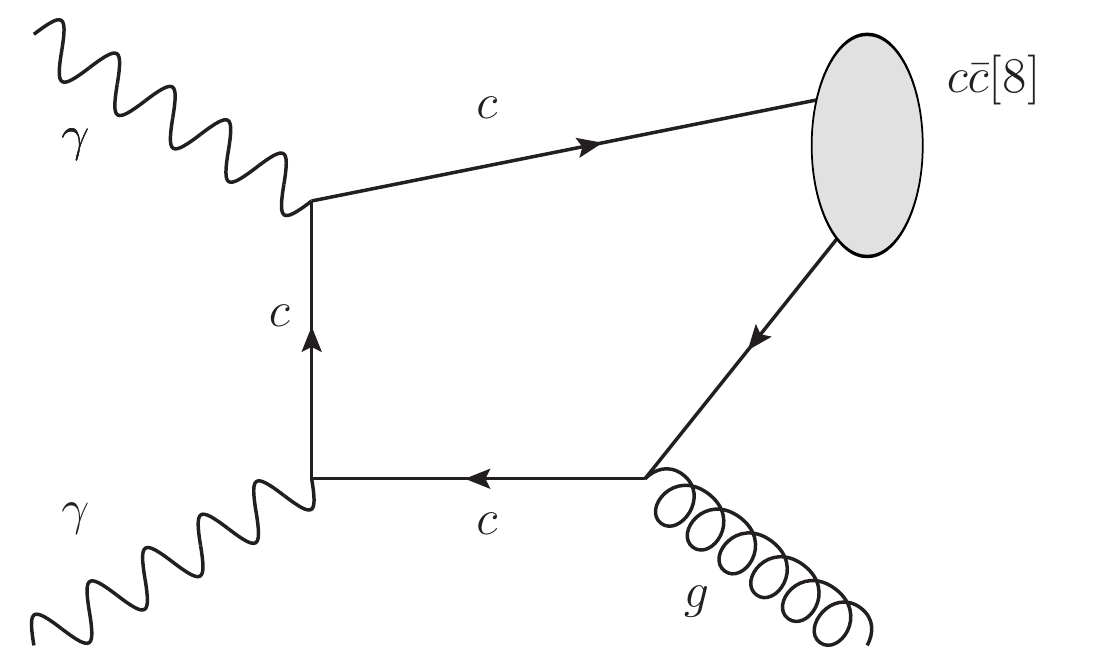}
	\includegraphics[width=0.24\textwidth]{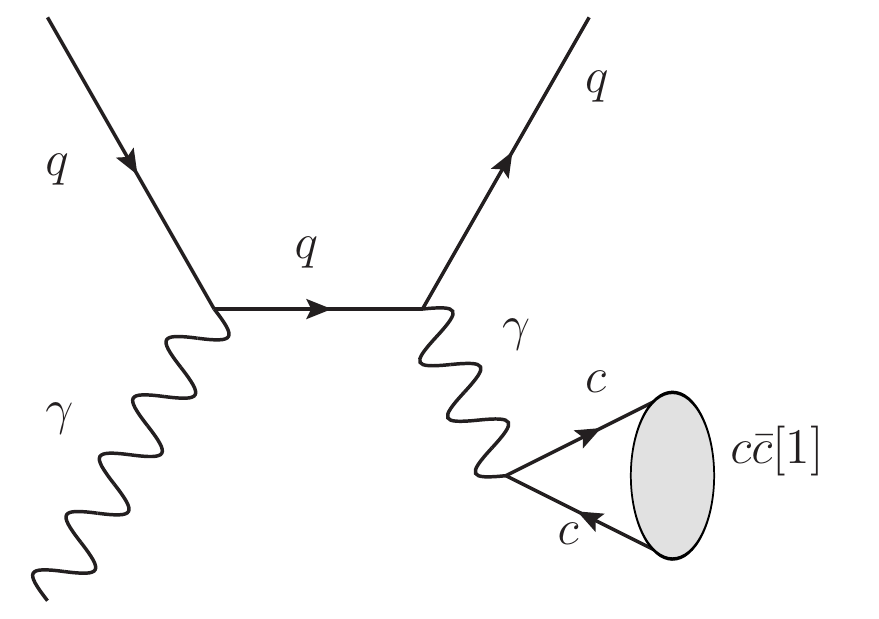}		\includegraphics[width=0.24\textwidth]{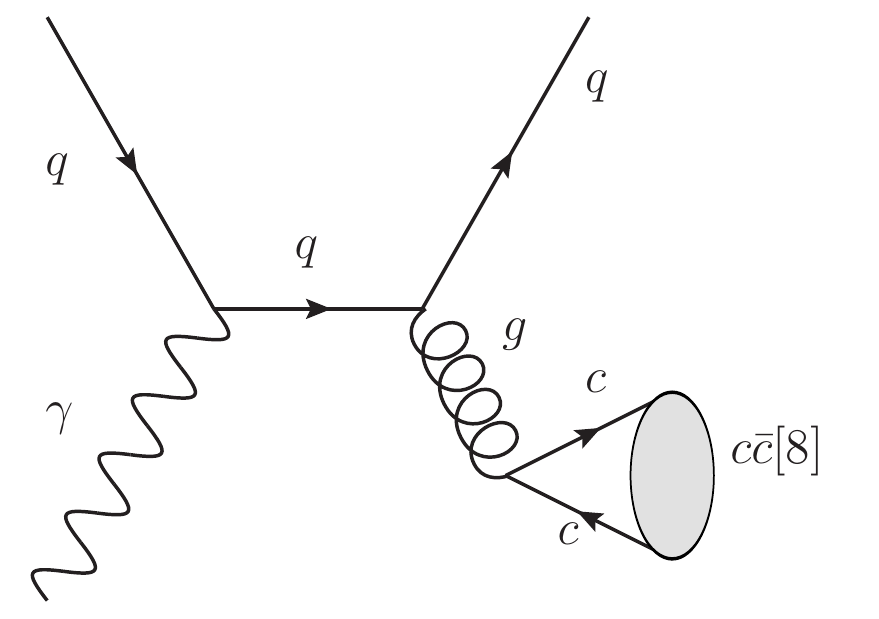}
	\includegraphics[width=0.24\textwidth]{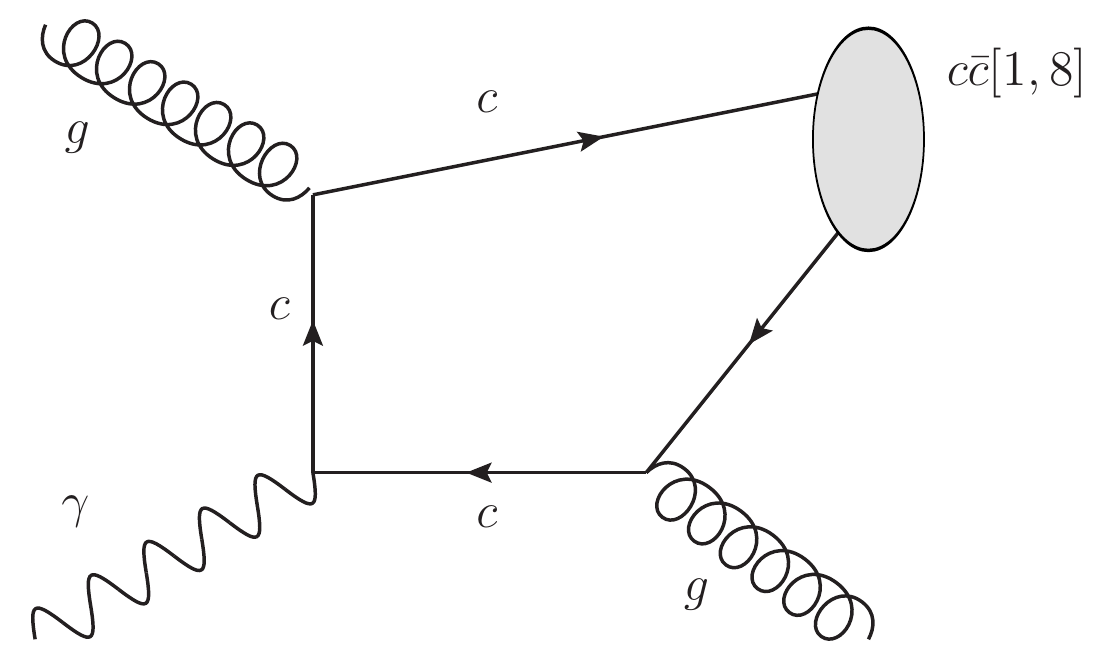}\\
	\includegraphics[width=0.24\textwidth]{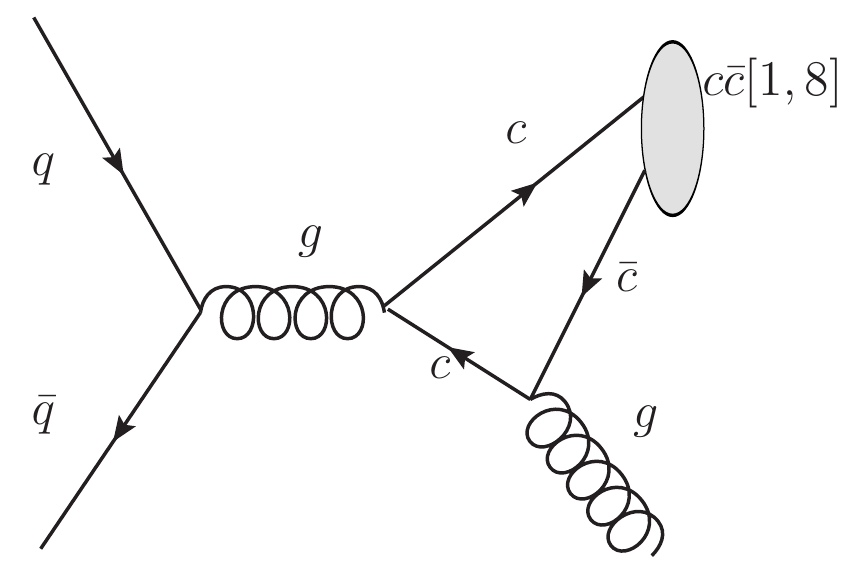}
	\includegraphics[width=0.24\textwidth]{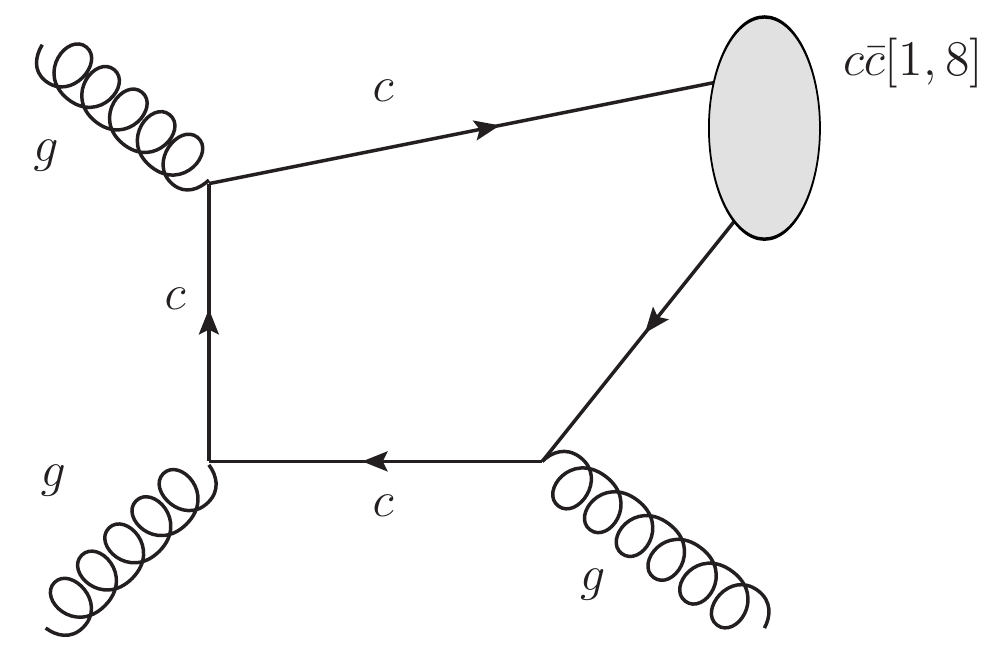}			\includegraphics[width=0.24\textwidth]{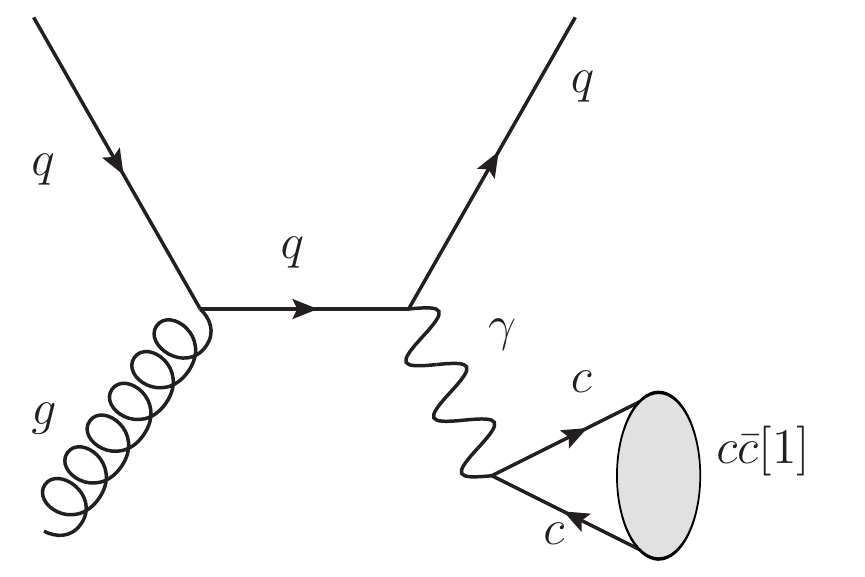}
	\includegraphics[width=0.24\textwidth]{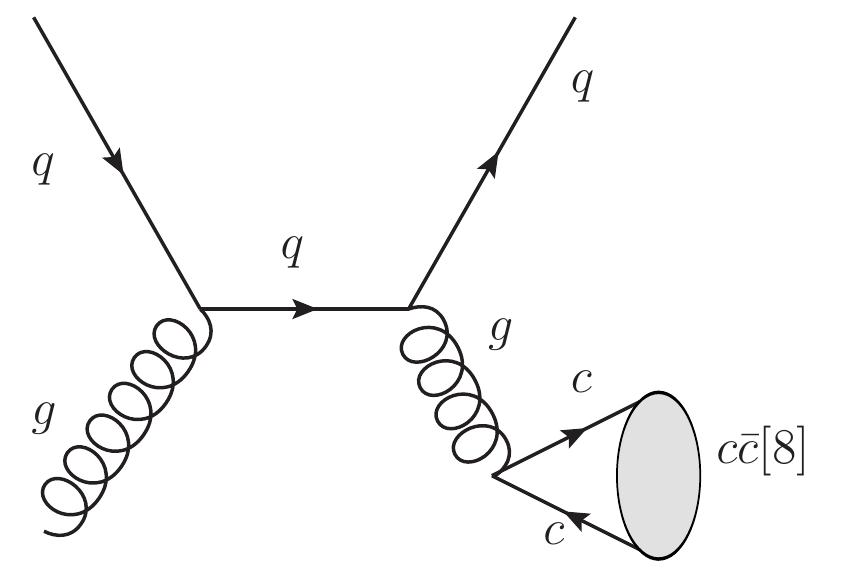}	
	\caption{Some Feynman diagrams of the photoproduction. The diagrams are drawn by JaxoDraw~\cite{Binosi:2003yf}.}
	\label{fig:21}
\end{figure}

Following are all the sub-processes to be calculated, which are for three production mechanisms. As for the direct photoproduction, we have
\begin{equation}
\begin{aligned}
&\gamma + \gamma \rightarrow c\bar{c}[{}^3\!S_1^{[\textbf{1}]},{}^1\!S_0^{[\textbf{8}]},{}^3\!S_1^{[\textbf{8}]},{}^3\!P_J^{[\textbf{8}]}] + c+\bar{c} \rightarrow J/\psi(\psi(2S)) +X,\\
&\gamma + \gamma \rightarrow c\bar{c}[{}^3\!P_J^{[\textbf{1}]},{}^3\!S_1^{[\textbf{8}]}] + c+\bar{c}\rightarrow \chi_{cJ} +X,\\
&\gamma + \gamma \rightarrow c\bar{c}[{}^3\!S_1^{[\textbf{8}]}] + g\rightarrow J/\psi(\psi(2S),\chi_{cJ}) +X.
\end{aligned}
\end{equation}
As for single resolved photoproduction, we have
\begin{equation}
\begin{aligned}
&\gamma + q(\bar{q},q=u,d,s) \rightarrow c\bar{c}[{}^3\!S_1^{[\textbf{1}]},{}^1\!S_0^{[\textbf{8}]},{}^3\!S_1^{[\textbf{8}]},{}^3\!P_J^{[\textbf{8}]}] + q(\bar{q})\rightarrow J/\psi(\psi(2S)) +X,\\
&\gamma + g \rightarrow c\bar{c}[{}^3\!S_1^{[\textbf{1}]},{}^1\!S_0^{[\textbf{8}]},{}^3\!S_1^{[\textbf{8}]},{}^3\!P_J^{[\textbf{8}]}] + c+\bar{c}\rightarrow J/\psi(\psi(2S)) +X,\\
&\gamma + g \rightarrow c\bar{c}[{}^3\!S_1^{[\textbf{1}]},{}^1\!S_0^{[\textbf{8}]},{}^3\!S_1^{[\textbf{8}]},{}^3\!P_J^{[\textbf{8}]}] + g\rightarrow J/\psi(\psi(2S)) +X,\\
&\gamma + q(\bar{q},q=u,d,s) \rightarrow c\bar{c}[{}^3\!P_J^{[\textbf{1}]},{}^3\!S_1^{[\textbf{8}]}] + q(\bar{q})\rightarrow \chi_{cJ} +X,\\
&\gamma + g \rightarrow c\bar{c}[{}^3\!P_J^{[\textbf{1}]},{}^3\!S_1^{[\textbf{8}]}] + c+\bar{c}\rightarrow \chi_{cJ} +X,\\
&\gamma + g \rightarrow c\bar{c}[{}^3\!S_1^{[\textbf{8}]}] + g\rightarrow \chi_{cJ} +X.
\end{aligned}
\end{equation}
As for double resolved photoproduction, we have
\begin{equation}
\begin{aligned}
&q(q=u,d,s) + \bar{q} \rightarrow c\bar{c}[{}^3\!S_1^{[\textbf{1}]},{}^1\!S_0^{[\textbf{8}]},{}^3\!S_1^{[\textbf{8}]},{}^3\!P_J^{[\textbf{8}]}] + g \rightarrow J/\psi(\psi(2S)) +X,\\
&q(q=u,d,s) + \bar{q} \rightarrow c\bar{c}[{}^3\!S_1^{[\textbf{1}]},{}^1\!S_0^{[\textbf{8}]},{}^3\!S_1^{[\textbf{8}]},{}^3\!P_J^{[\textbf{8}]}] + c+\bar{c} \rightarrow J/\psi(\psi(2S)) +X,\\
&g + g \rightarrow c\bar{c}[{}^3\!S_1^{[\textbf{1}]},{}^1\!S_0^{[\textbf{8}]},{}^3\!S_1^{[\textbf{8}]},{}^3\!P_J^{[\textbf{8}]}] + g \rightarrow J/\psi(\psi(2S)) +X,\\
&g + g \rightarrow c\bar{c}[{}^3\!S_1^{[\textbf{1}]},{}^1\!S_0^{[\textbf{8}]},{}^3\!S_1^{[\textbf{8}]},{}^3\!P_J^{[\textbf{8}]}] + c+\bar{c} \rightarrow J/\psi(\psi(2S)) +X,\\
&g + q(\bar{q},q=u,d,s) \rightarrow c\bar{c}[{}^3\!S_1^{[\textbf{1}]},{}^1\!S_0^{[\textbf{8}]},{}^3\!S_1^{[\textbf{8}]},{}^3\!P_J^{[\textbf{8}]}] + q(\bar{q})\rightarrow J/\psi(\psi(2S)) +X,\\
&q(q=u,d,s) + \bar{q} \rightarrow c\bar{c}[{}^3\!P_J^{[\textbf{1}]},{}^3\!S_1^{[\textbf{8}]}] + g \rightarrow \chi_{cJ} +X,\\
&q(q=u,d,s) + \bar{q} \rightarrow c\bar{c}[{}^3\!P_J^{[\textbf{1}]},{}^3\!S_1^{[\textbf{8}]}] + c+\bar{c} \rightarrow \chi_{cJ} +X,\\
&g + g \rightarrow c\bar{c}[{}^3\!P_J^{[\textbf{1}]},{}^3\!S_1^{[\textbf{8}]}] + g \rightarrow \chi_{cJ} +X,\\
&g + g \rightarrow c\bar{c}[{}^3\!P_J^{[\textbf{1}]},{}^3\!S_1^{[\textbf{8}]}] + c+\bar{c} \rightarrow \chi_{cJ} +X,\\
&g + q(\bar{q},q=u,d,s) \rightarrow c\bar{c}[{}^3\!P_J^{[\textbf{1}]},{}^3\!S_1^{[\textbf{8}]}] + q(\bar{q})\rightarrow \chi_{cJ} +X.\\
\end{aligned}
\end{equation}

Some Feynman diagrams of these photoproduction processes are presented in figure~\ref{fig:21}. The well-established package, Feynman Diagram Calculation (FDC)~\cite{Wang:2004du}, is used to do the analytical and numerical calculations. In FDC, the standard projection method~\cite{Bodwin:2002cfe} is employed to deal with the hard process. After dealing with the squared amplitudes analytically, FDC generates FORTRAN codes for numerical integration of phase space.

\section{Numerical results and discussions}
\label{sec:3}

To do the numerical calculation, we choose the electromagnetic fine structure constant $\alpha=1/137$ and the one-loop running strong coupling constant $\alpha_s(\mu_r)$.
To conserve the gauge invariant of the hard scattering amplitude, the charm quark mass,  $m_c$, is set approximately as $m_c=m_H/2$, where the charmonia masses $m_H= 3.097, 3.415, 3.511, 3.556, 3.686 \mathrm{~GeV}$~\cite{Tanabashi:2018oca} for $H=J / \psi, \chi_{c J}(J=0,1,2)$ and $\psi(2 S)$, respectively. The branching ratios are taken as $Br(\psi(2 S) \rightarrow J / \psi)=0.61$ and $Br\left(\chi_{c J} \rightarrow J / \psi\right)=0.014,0.343,0.19$ for $J=0,1,2$~\cite{Tanabashi:2018oca}.
In dealing with the feed-down contributions, a shift of the transverse momentum of charmonium, $p_{t}^{H} \approx p_{t}^{H^{\prime}} \times\left(m_{H} / m_{H^{\prime}}\right)$, is used. The renormalization scale $\mu_r$ is set to be $\mu_r=m_T=\sqrt{m_{H}^{2}+(p_{t}^H)^{2}}$. Taking $\mu_r=m_T/2$, the cross section ($\sqrt{s}=500\mathrm{~GeV}$) shall be increased by about $80\%$, and taking $\mu_r=2m_T$, the cross section shall decreased by about $40\%$.
Such a large scale dependence could be tamed by higher order calculation or a proper scale setting, c.f. Ref.\cite{Wu:2019mky}. As for the non-perturbative LDMEs, we take~\cite{Feng:2018ukp},
\begin{equation}
\begin{aligned}
\langle\mathcal{O}^{\psi}({ }^{3} S_{1}^{[\textbf{1}]})\rangle &=\frac{3 N_{c}}{2 \pi}|R_{\psi}(0)|^{2}, \\
\langle{O}^{\chi_{cJ}}({ }^{3} P_{J}^{[\textbf{1}]})\rangle &=\frac{3}{4 \pi}(2 J+1)|R_{\chi_{c}}^{\prime}(0)|^{2},\\
\langle O^{J / \psi}({ }^{1} S_{0}^{[\textbf{8}]})\rangle &=5.66\times 10^{-2} \mathrm{~GeV}^{3}, \\
\langle O^{J / \psi}({ }^{3} S_{1}^{[\textbf{8}]})\rangle &=1.77 \times 10^{-2} \mathrm{~GeV}^{3}, \\
{\langle O^{J / \psi}({ }^{3} P_{0}^{[\textbf{8}]})\rangle}/{m_{c}^{2}} &=3.42 \times 10^{-3} \mathrm{~GeV}^{3},\\
\langle O^{\psi(2S)}({ }^{1} S_{0}^{[\textbf{8}]})\rangle &=-1.20 \times 10^{-4} \mathrm{~GeV}^{3}, \\
\langle O^{\psi(2S)}({ }^{3} S_{1}^{[\textbf{8}]})\rangle &=3.40 \times 10^{-3} \mathrm{~GeV}^{3}, \\
{\langle O^{\psi(2S)}({ }^{3} P_{0}^{[\textbf{8}]})\rangle}/{m_{c}^{2}} &=4.20 \times 10^{-3} \mathrm{~GeV}^{3},\\
\langle O^{\chi_{c0}}({ }^{3} S_{1}^{[\textbf{8}]})\rangle &=2.21 \times 10^{-3} \mathrm{~GeV}^{3},
\end{aligned}
\end{equation}
where the wave functions at the origin are given as $|R_{J / \psi}(0)|^{2}=0.81 \mathrm{~GeV}^{3}$, $|R_{\psi(2 S)}(0)|^{2}=0.53 \mathrm{~GeV}^{3}$ and $|R_{\chi_{c}}^{\prime}(0)|^{2}=0.075 \mathrm{~GeV}^{5}$~\cite{Eichten:1995ch}.

\begin{table}[tp]
	\centering
	\begin{tabular}{|c|cccc|}
		\hline
		$\sqrt{S}(\mathrm{GeV})$ & $\sigma_{directJ/\psi}$ & $\sigma_{\psi(2S)\rightarrow J/\psi}$ & $\sigma_{\chi_{cJ}\rightarrow J/\psi}$ & $\sigma_{promptJ/\psi}$\\
		\hline
		250 & $420.13$ & $28.27$ & $2.21$ & $450.61$ \\
		500 & $667.84$ & $45.18$ & $4.44$ & $717.46$ \\
		1000 & $1036.89$ & $70.66$ & $8.44$ & $1115.99$ \\
		\hline
	\end{tabular}
	\caption{\label{tab:cross-section1}The integrated cross sections (in unit of pb) for prompt $J/\psi$ photoproduction under different collision energies at the ILC. The cut $p_t>1$ is set for $J/\psi$. Both the CS and the CO channels have been summed up.}
\end{table}

\begin{table}[tp]
	\centering
	\begin{tabular}{|c|cccccc|}
		\hline
		$\sqrt{S}(\mathrm{GeV})$ & $\sigma^{directJ/\psi}_{{}^{3}S_{1}^{[\textbf{1}]}}$ & $\sigma^{\psi(2S)\rightarrow J/\psi}_{{}^{3}S_{1}^{[\textbf{1}]}}$
		& $\sigma^{\chi_{c J}\rightarrow J/\psi}_{{}^{3}P_{J}^{[\textbf{1}]}}$
		& $\sigma^{directJ/\psi}_{{}^{1}S_{0}^{[\textbf{8}]}}$ & $\sigma^{directJ/\psi}_{{}^{3}S_{1}^{[\textbf{8}]}}$ & $\sigma^{directJ/\psi}_{{}^{3}P_{J}^{[\textbf{8}]}}$\\
		\hline
		250 & $21.61$ & $6.89$ & $2.05$ & $348.22$ & $0.34$ & $49.95$\\
		500 & $32.12$ & $10.44$ & $4.15$ & $555.48$ & $0.61$ & $79.63$ \\
		1000 & $47.34$ & $15.60$ & $7.92$ & $864.47$ & $1.10$ & $123.99$ \\
		\hline
	\end{tabular}
	\caption{\label{tab:cross-section2}The integrated cross sections (in unit of pb) from various channels for $J/\psi$ photoproduction under different collision energies at the ILC. The cut $p_t>1$ is set for $J/\psi$.}
\end{table}

In table \ref{tab:cross-section1}, the  integrated cross sections of prompt $J/\psi$ photoproduction at the ILC under different energies are listed. It can be seen that the integrated cross section becomes larger with the increment of the collision energy, while the ratios of feed-down contributions are not sensitive to the energy, which are about $7\%$.
The contribution of direct $J/\psi$ photoproduction dominates over those from the feed-down. Due to the large cross section, a huge number of $J/\psi$ events are expected to be generated via the photoproduction at the ILC.

Table \ref{tab:cross-section2} presents the integrated cross sections for different channels of $J/\psi$ photoproduction.
The color-octet channels are dominant and two channels, ${}^{1}S_{0}^{[\textbf{8}]}$ and ${}^{3}P_{J}^{[\textbf{8}]}$, provide about $95\%$ contributions to the direct $J/\psi$ production.
For the $J/\psi$ photoproduction at the ILC, the NRQCD factorization framework and the CS model therefore give predictions that differ by one order of magnitude.
In the color-singlet channel, the feed-down contributions from $\psi(2S)$ and $\chi_{c J}$ are significant, where the situation is very different from that in the color-octet channels.
Although the integrated cross section for the CS channel is very small when compared with that of the CO,  it itself is still a sizable cross section.
For example, if setting the integrated luminosity of the ILC as $\mathcal{O}(10^4)\mathrm{~fb^{-1}}$, there would be $\mathcal{O}(10^8)$ $J/\psi$ mesons to be produced via only the CS channels.
Consequently, the measurement of $J/\psi$ photoproduction at the ILC could be done precisely and it shall be very helpful to test NRQCD factorization and to further study physics of heavy quarkonium.
In the following we take $\sqrt{S}=500\mathrm{~GeV}$ for more discussions.

\begin{figure}[t]
	\centering % \begin{center}/\end{center} takes some additional
	\includegraphics[width=.5\textwidth]{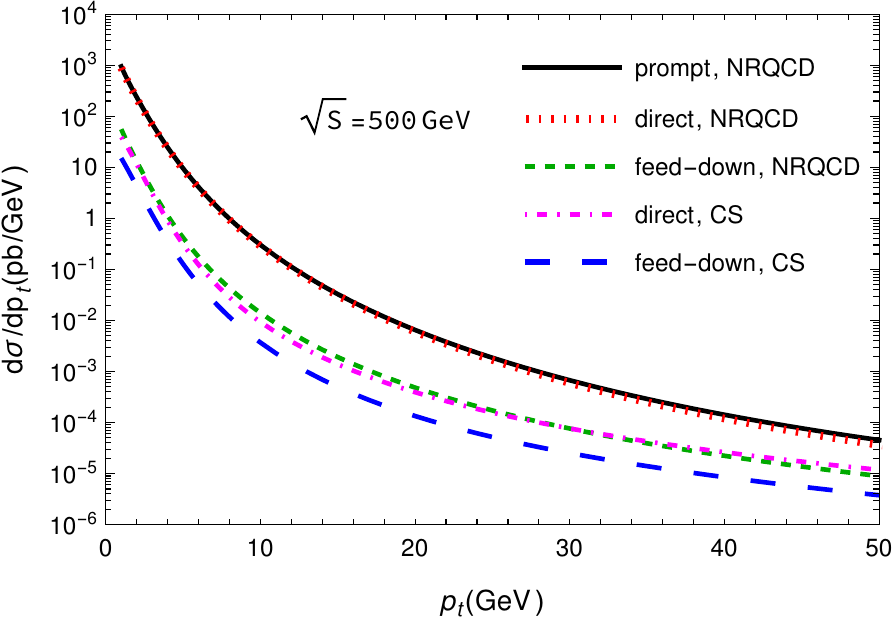}
	\caption{\label{fig:31} The $p_t$ distributions for prompt $J/\psi$ photoproduction at the ILC ($\sqrt{S}=500\mathrm{~GeV}$).}
\end{figure}

Figure~\ref{fig:31} shows the prompt $J/\psi$ photoproduction in terms of transverse momentum distributions, where the direct and feed-down channels are displayed separately, both for the NRQCD and the CSM predictions. We can see that the direct production dominate in the whole $p_t$ region.

\begin{figure}[t]
	\centering % \begin{center}/\end{center} takes some additional
	\includegraphics[width=.32\textwidth]{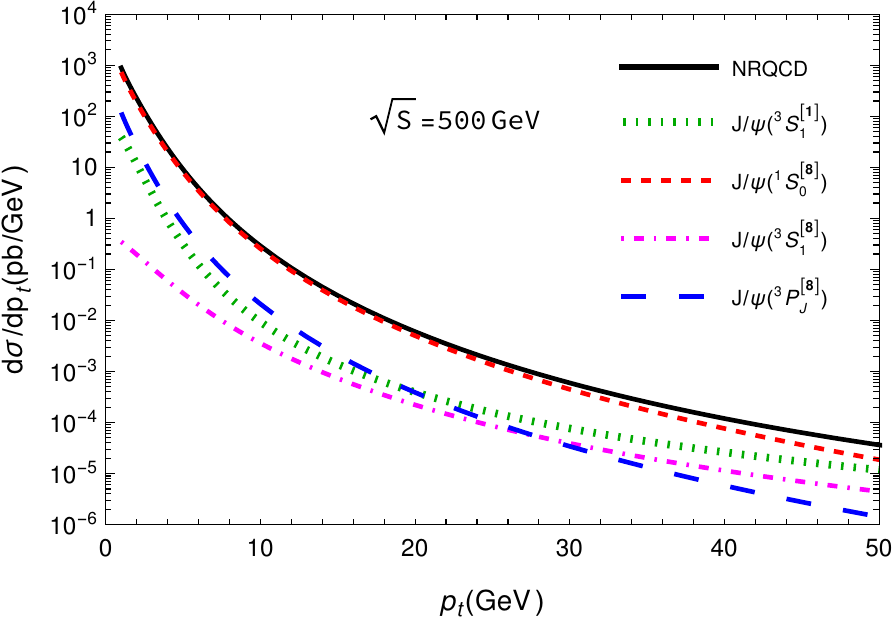}
	\includegraphics[width=.32\textwidth]{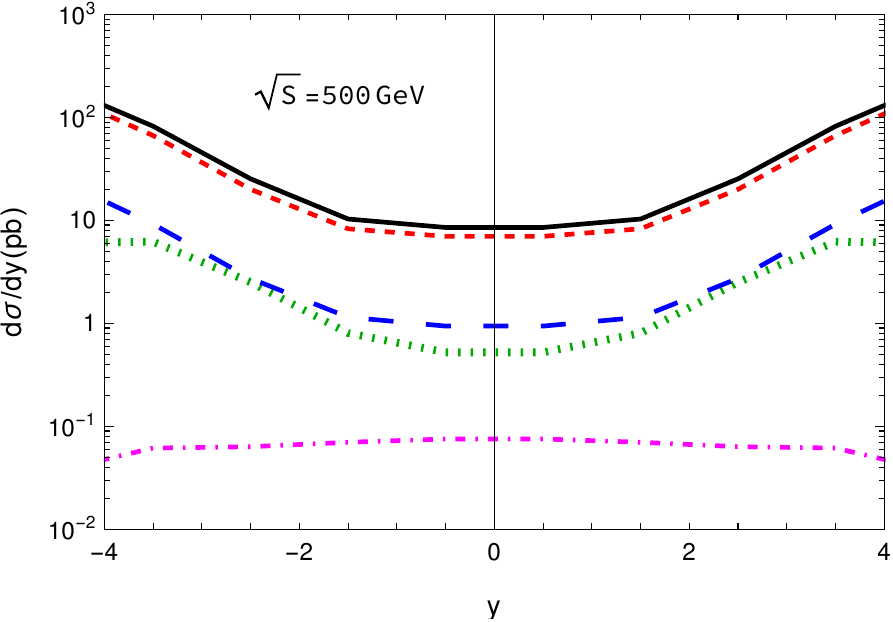}
	\includegraphics[width=.32\textwidth]{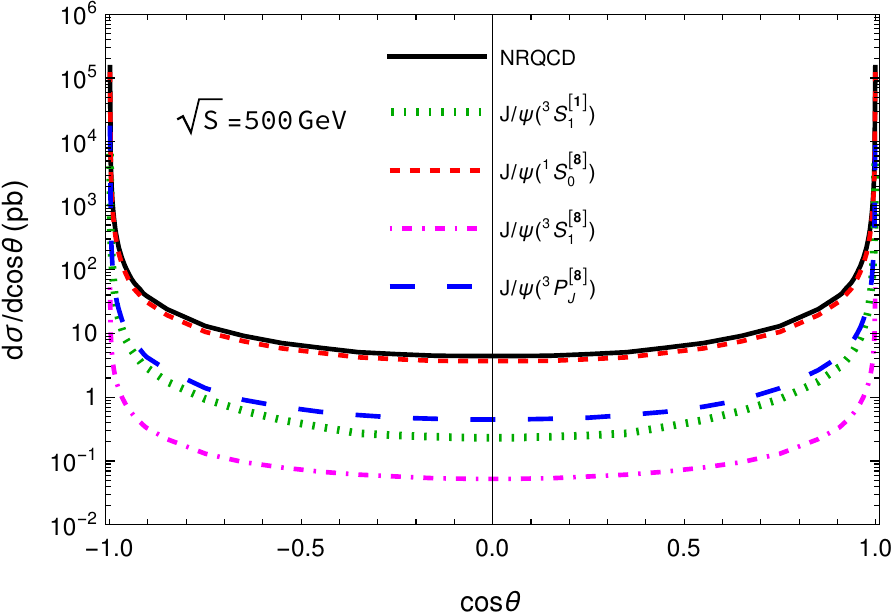}	
	\caption{\label{fig:32} Kinematic distributions for different intermediate $c\bar{c}$ states of $J/\psi$ photoproduction at the ILC ($\sqrt{S}=500\mathrm{~GeV}$). The $y$ and $\cos\theta$ distributions are plotted under the cut $p_t>1$. $y$ curves use same legends as those of $p_t$.}
\end{figure}

\begin{figure}[t]
	\centering
	\includegraphics[width=.32\textwidth]{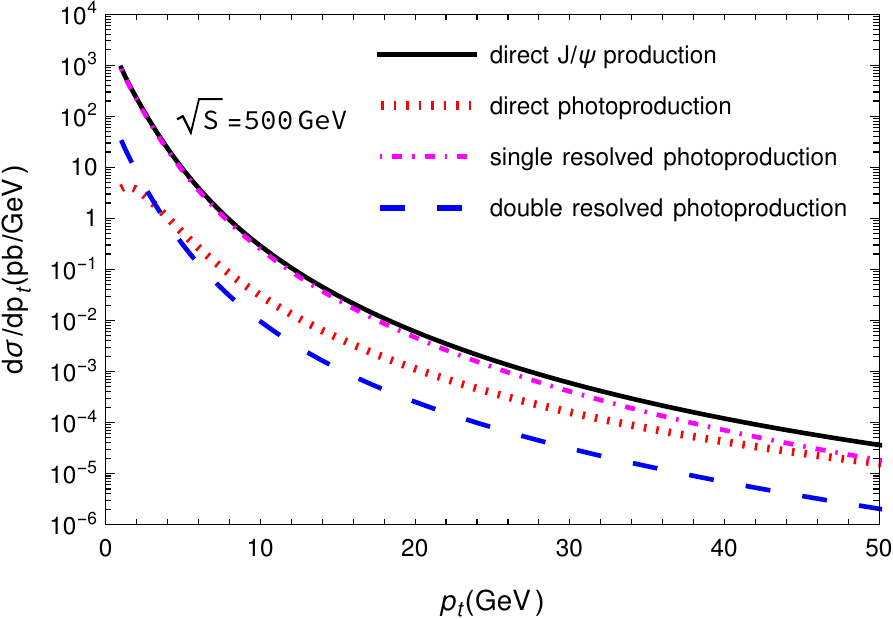}
	\includegraphics[width=.32\textwidth]{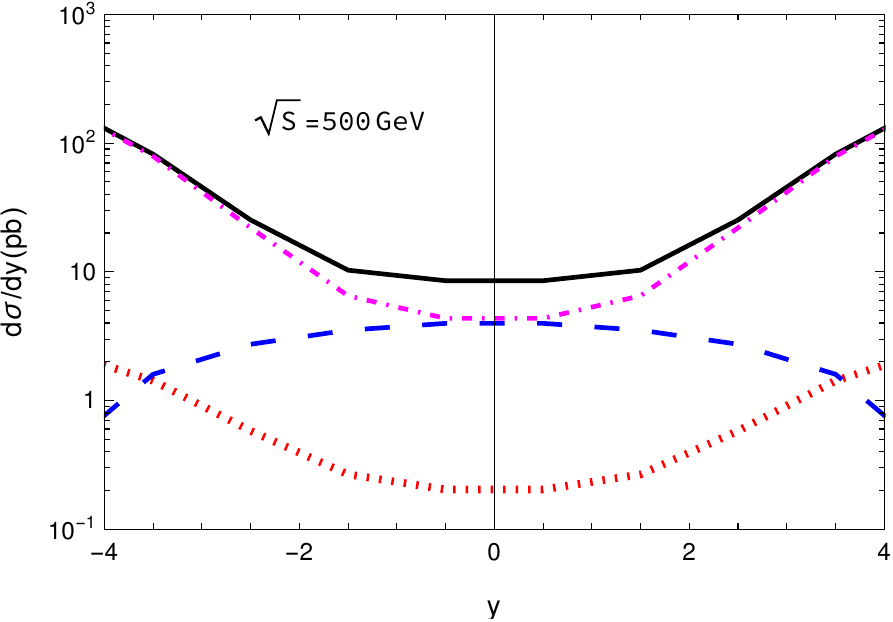}
	\includegraphics[width=.32\textwidth]{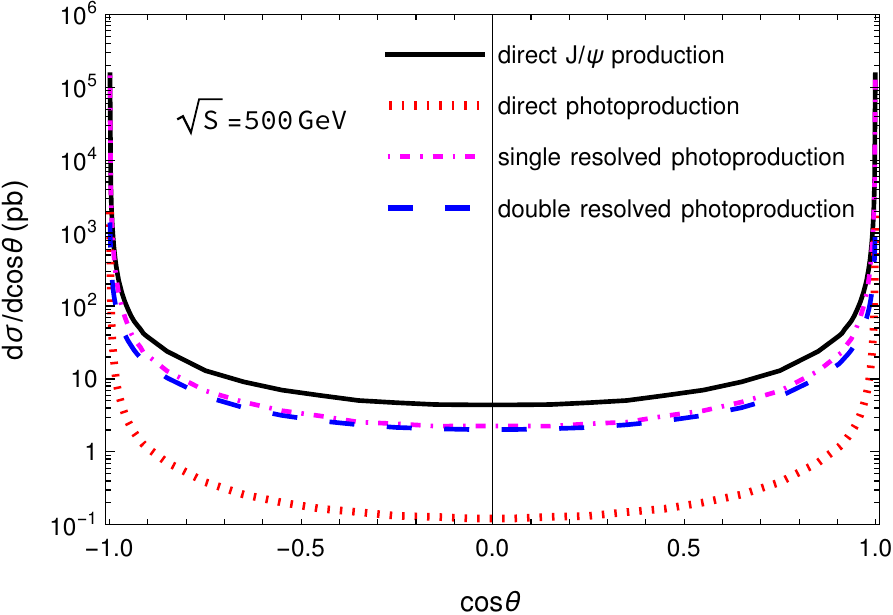}
	\caption{\label{fig:33} Kinematic distributions for the resolved photoproduction of $J/\psi$ at the ILC ($\sqrt{S}=$
	$500$ $\mathrm{GeV}$). The $y$ and $\cos\theta$ distributions are plotted under the cut $p_t>1$. All the distributions are plotted for the NRQCD predictions and $y$ curves use same legends as those of $p_t$.}
\end{figure}
Figure~\ref{fig:32} shows the kinematic distributions for different intermediate $c\bar{c}$ Fock states.
Here three kind of kinematic distributions of transverse momentum($p_t$), rapidity($y$) and angular
($\cos\theta$), are presented.
$\theta$ is the angle between $J/\psi$ and the collision beams.
It can be seen that these channels have different $p_t$ behaviors.
The CO channels dominate in small $p_t$ regions while the contribution from the CS channel become relatively important in large $p_t$ region.
In the rapidity and angular distributions, the curves of different channels do not intersect with each other in the whole region, and the ${}^{1}S_{0}^{[\textbf{8}]}$ channel is always primarily dominant.
The curves of these two kinematic distributions change gently in the wide middle regions,
which means there would be enough events in the whole region to make well measurement.
Taking the integrated luminosity of the ILC to be $\mathcal{O}(10^4)\mathrm{~fb^{-4}}$ as reference,
in the lowest region of $y$ and $\cos\theta$, e.g., $-0.5<y<0.5$ and $-0.05<\cos\theta<0.05$, there would produce respectively about $10^5$ and $10^4$ $J/\psi$ mesons by the NRQCD prediction,
and about $10^4$ and $10^3$ $J/\psi$ by the CS model.
Due to the high luminosity and the large cross section of $J/\psi$ photoproduction, precise measurements of these kinematic distributions at the ILC are very promising.
These kinematic distributions can be used to contract and constrain the LDMEs accurately.

In figure~\ref{fig:33}, kinematic distributions of the direct photoproduction, the single resolved photoproduction and the double resolved photoproduction are presented.
In the region of $1\mathrm{~GeV}<p_t<50\mathrm{~GeV}$, the single resolved photoproduction contributes $95\%$ the integrated cross section. Although in large $p_t$ region, the direct photoproduction give more contributions than the single resolved photoproduction, there may be not enough events to make precise measurement.
In the region of $\cos\theta$, about $-0.9<\cos\theta<0.9$, the single resolved and double resolved channels provide contributions of the same magnitude, while the direct production is always negligible.

\begin{figure}[t]
	\centering
	\includegraphics[width=.36\textwidth]{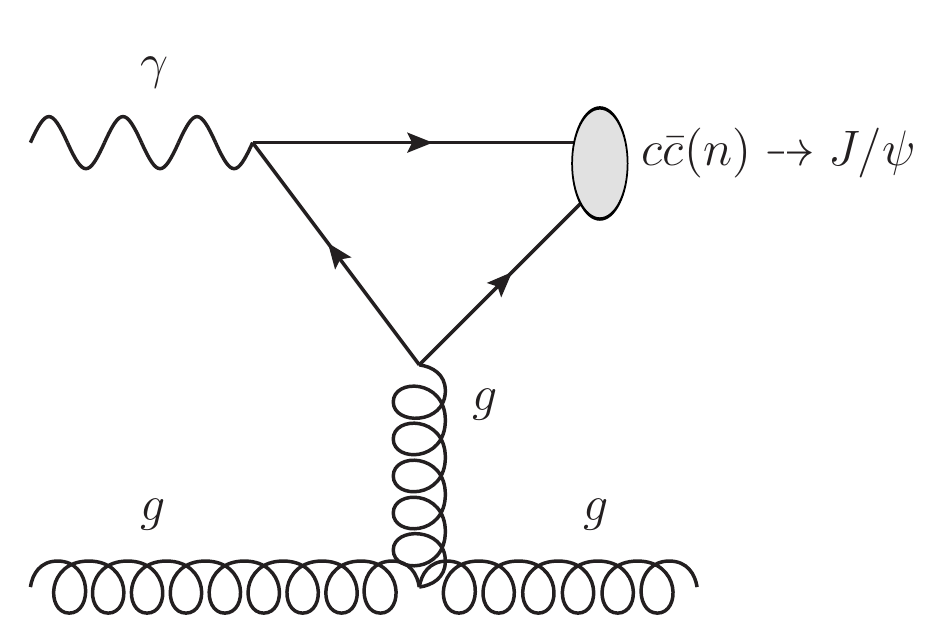}
	\caption{\label{fig:34} Feynman diagram of single resolved photoproduction $\gamma+g\rightarrow c\bar{c}(n={}^{1}S_{0}^{[\textbf{8}]},$ ${}^{3}P_{J=0,1,2}^{[\textbf{8}]})+g \rightarrow J/\psi+X$  .}
\end{figure}

The single resolved sub-processes $\gamma+g\rightarrow c\bar{c}({}^{1}S_{0}^{[\textbf{8}]},{}^{3}P_{J=0,1,2}^{[\textbf{8}]})+g \rightarrow J/\psi+X$ provide the most contributions for $J/\psi$ photoproduction at the ILC. This can be explained by looking insight into their Feynman diagrams.
The diagrams shown in figure~\ref{fig:34} are absent for other sub-processes due to parity and color conservation, and they provide the main contributions of ${}^{1}S_{0}^{[\textbf{8}]},{}^{3}P_{J}^{[\textbf{8}]}$ channels.
The squared invariant mass of the gluon propagator in figure~\ref{fig:34} can be expressed as,
\begin{equation}
\begin{aligned}
k^{2} &=4 m_{c}^{2}-x \sqrt{S} M_{t} e^{-y} \\
&=4 m_{c}^{2}-2\left(E_{J/\psi}+E_{g}\right) M_{t} e^{-y} \\
&=-4 m_{c}^{2} e^{-2 y}-\left(1+e^{-2 y}\right)\left(p_{t}^{J/\psi}\right)^{2}-2 E_{g} M_{t} e^{-y},
\end{aligned}
\end{equation}
from which it can be seen that $1/k^2$ and hence the cross section could be very large in small $p_t$ and large $y$ regions.
This feature is also illustrated in the $p_t$ and $y$ distributions of figure~\ref{fig:32}.

\begin{figure}[t]
	\centering % \begin{center}/\end{center} takes some additional
	\includegraphics[width=.32\textwidth]{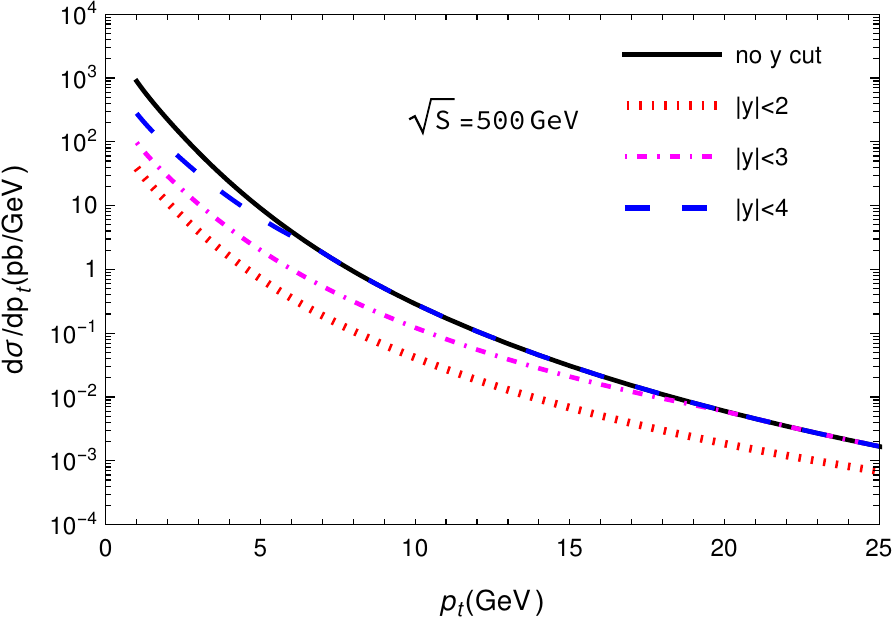}
	\includegraphics[width=.32\textwidth]{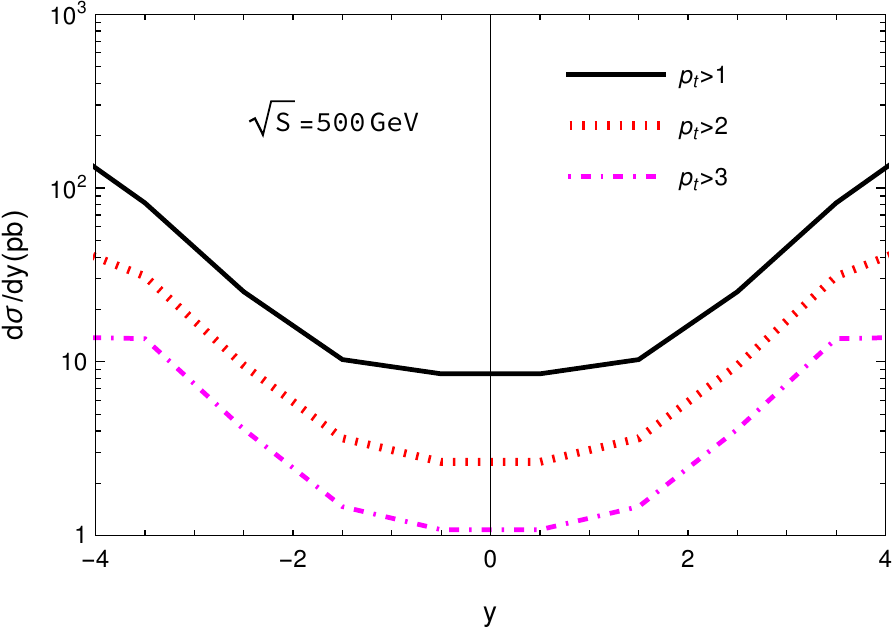}
	\includegraphics[width=.32\textwidth]{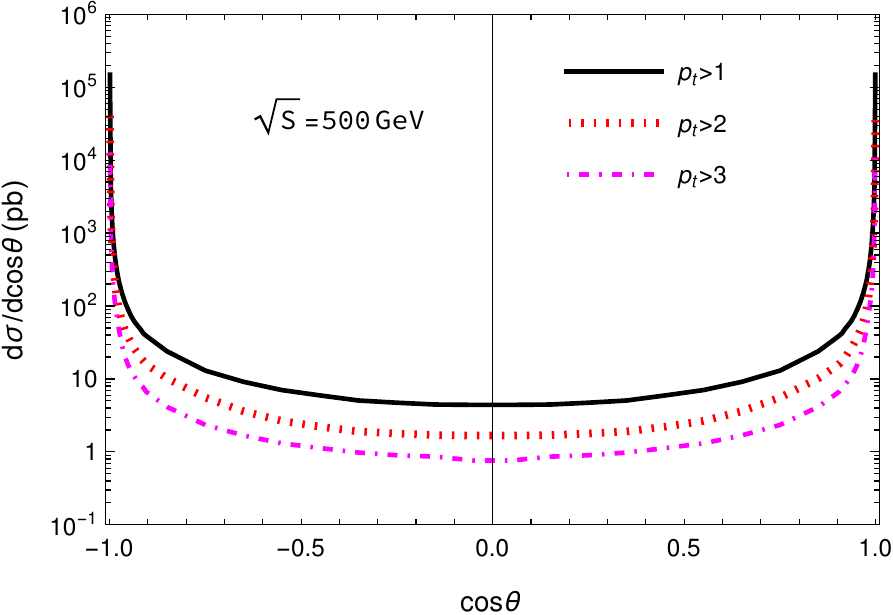}
	\caption{\label{fig:35} Kinematic distributions under various cuts for $J/\psi$ photoproduction at the ILC($\sqrt{S}=500\mathrm{~GeV}$). All the distributions are plotted for the NRQCD predictions.}
\end{figure}

To make our predictions more referential, we also calculate the integrated cross sections under various kinematic cuts, listed in table~\ref{tab:cross-section3}, and give kinematic distributions under the cuts, shown in figure~\ref{fig:35}.
It can be seen that these cuts have great effect on the cross sections and the kinematic distributions. After imposed these cuts, however, the cross section and thus the number of $J/\psi$ meson to be produced are still sizable.

\begin{table}[t]
	\centering
	\begin{tabular}{|c|ccc|}
		\hline
		cuts & $p_t>1\mathrm{~GeV}$ & $p_t>2\mathrm{~GeV}$
		& $p_t>3\mathrm{~GeV}$\\
		\hline
		$|y|<2$ & $38.85$ & $13.94$ & $5.84$ \\
		$|y|<3$ & $97.47$ & $35.83$ & $15.29$ \\
		$|y|<4$ & $278.09$ & $103.62$ & $44.61$ \\
		\hline
	\end{tabular}
	\caption{\label{tab:cross-section3} The integrated cross sections (in unit of pb) under various cuts for $J/\psi$ photoproduction at the ILC ($\sqrt{S}=500\mathrm{~GeV}$). All the predictions are based on NRQCD.}
\end{table}

\section{Summary}
\label{sec:4}

In this work, inclusive $J/\psi$ photoproduction at the future ILC is preliminarily studied at the leading order of $\alpha_s$ within the NRQCD factorization framework.
Both the color-octet channels and the resolved photoproduction processes are considered.
The numerical results show that the color-octet channels and the single resolved processes dominate primarily the production.
Due to the large cross section of the photoproduction and high luminosity of the ILC, it could be expected that there will be numerous $J/\psi$ mesons to be produced, and various kinematic distributions will be measured precisely.
Consequently, the photoproduction of $J/\psi$ at the ILC provides a good laboratory to test the NRQCD factorization framework and to deep our knowledge of the production mechanism of heavy quarkonium.

\acknowledgments

This work was supported in part by the Natural Science Foundation of China under Grants No. 12147116, No. 12175025, No. 12005028 and No. 12147102, by the China Postdoctoral Science Foundation under Grant No. 2021M693743 and by the Fundamental Research Funds for the Central Universities under Grant No. 2020CQJQY-Z003.

%\bibliographystyle{unsrt}
%\bibliography{refs}

\begin{thebibliography}{10}
	
	\bibitem{Bodwin:1994jh}
	Geoffrey~T. Bodwin, Eric Braaten, and G.~Peter Lepage.
	\newblock {Rigorous QCD analysis of inclusive annihilation and production of
		heavy quarkonium}.
	\newblock {\em Phys. Rev.}, D51:1125--1171, 1995.
	\newblock [Erratum: Phys. Rev.D55,5853(1997)].
	
	\bibitem{Berger:1980ni}
	Edmond~L. Berger and Daniel~L. Jones.
	\newblock {Inelastic Photoproduction of J/psi and Upsilon by Gluons}.
	\newblock {\em Phys. Rev. D}, 23:1521--1530, 1981.
	
	\bibitem{Baier:1981uk}
	R.~Baier and R.~Ruckl.
	\newblock {Hadronic Production of J/psi and Upsilon: Transverse Momentum
		Distributions}.
	\newblock {\em Phys. Lett. B}, 102:364--370, 1981.
	
	\bibitem{Humpert:1986cy}
	B.~Humpert.
	\newblock {NARROW HEAVY RESONANCE PRODUCTION BY GLUONS}.
	\newblock {\em Phys. Lett. B}, 184:105--107, 1987.
	
	\bibitem{Gastmans:1986qv}
	R.~Gastmans, W.~Troost, and Tai~Tsun Wu.
	\newblock {Cross-Sections for Gluon + Gluon ---\ensuremath{>} Heavy Quarkonium
		+ Gluon}.
	\newblock {\em Phys. Lett. B}, 184:257--260, 1987.
	
	\bibitem{Gastmans:1987be}
	R.~Gastmans, W.~Troost, and Tai~Tsun Wu.
	\newblock {Production of Heavy Quarkonia From Gluons}.
	\newblock {\em Nucl. Phys. B}, 291:731, 1987.
	
	\bibitem{Campbell:2007ws}
	John~M. Campbell, F.~Maltoni, and F.~Tramontano.
	\newblock {QCD corrections to J/psi and Upsilon production at hadron
		colliders}.
	\newblock {\em Phys. Rev. Lett.}, 98:252002, 2007.
	
	\bibitem{Gong:2008ft}
	Bin Gong, Xue~Qian Li, and Jian-Xiong Wang.
	\newblock {QCD corrections to $J/\psi$ production via color octet states at
		Tevatron and LHC}.
	\newblock {\em Phys. Lett. B}, 673:197--200, 2009.
	\newblock [Erratum: Phys.Lett.B 693, 612--613 (2010)].
	
	\bibitem{Butenschoen:2010rq}
	Mathias Butenschoen and Bernd~A. Kniehl.
	\newblock {Reconciling J/psi production at HERA, RHIC, Tevatron, and LHC with
		NRQCD factorization at next-to-leading order}.
	\newblock {\em Phys. Rev. Lett.}, 106:022003, 2011.
	
	\bibitem{Ma:2010yw}
	Yan-Qing Ma, Kai Wang, and Kuang-Ta Chao.
	\newblock {$J/psi (psi')$ production at the Tevatron and LHC at
		O($\alpha_s^4v^4$) in nonrelativistic QCD}.
	\newblock {\em Phys. Rev. Lett.}, 106:042002, 2011.
	
	\bibitem{Gong:2010bk}
	Bin Gong, Jian-Xiong Wang, and Hong-Fei Zhang.
	\newblock {QCD corrections to $\Upsilon$ production via color-octet states at
		the Tevatron and LHC}.
	\newblock {\em Phys. Rev. D}, 83:114021, 2011.
	
	\bibitem{Butenschoen:2012px}
	Mathias Butenschoen and Bernd~A. Kniehl.
	\newblock {J/psi polarization at Tevatron and LHC: Nonrelativistic-QCD
		factorization at the crossroads}.
	\newblock {\em Phys. Rev. Lett.}, 108:172002, 2012.
	
	\bibitem{Chao:2012iv}
	Kuang-Ta Chao, Yan-Qing Ma, Hua-Sheng Shao, Kai Wang, and Yu-Jie Zhang.
	\newblock {$J/\psi$ Polarization at Hadron Colliders in Nonrelativistic QCD}.
	\newblock {\em Phys. Rev. Lett.}, 108:242004, 2012.
	
	\bibitem{Gong:2012ug}
	Bin Gong, Lu-Ping Wan, Jian-Xiong Wang, and Hong-Fei Zhang.
	\newblock {Polarization for Prompt J/\ensuremath{\psi} and
		\ensuremath{\psi}(2s) Production at the Tevatron and LHC}.
	\newblock {\em Phys. Rev. Lett.}, 110(4):042002, 2013.
	
	\bibitem{Feng:2018ukp}
	Yu~Feng, Bin Gong, Chao-Hsi Chang, and Jian-Xiong Wang.
	\newblock {Remaining parts of the long-standing $J/\psi$ polarization puzzle}.
	\newblock {\em Phys. Rev. D}, 99(1):014044, 2019.
	
	\bibitem{CEPCStudyGroup:2018rmc}
	CEPC~Study Group.
	\newblock {CEPC Conceptual Design Report: Volume 1 - Accelerator}.
	\newblock 9 2018.
	
	\bibitem{CEPCStudyGroup:2018ghi}
	Mingyi Dong et~al.
	\newblock {CEPC Conceptual Design Report: Volume 2 - Physics \& Detector}.
	\newblock 2018.
	
	\bibitem{FCC:2018evy}
	A.~Abada et~al.
	\newblock {FCC-ee: The Lepton Collider}: {Future Circular Collider Conceptual
		Design Report Volume 2}.
	\newblock {\em Eur. Phys. J. ST}, 228(2):261--623, 2019.
	
	\bibitem{ILC:2007bjz}
	Gerald Aarons et~al.
	\newblock {International Linear Collider Reference Design Report Volume 2:
		Physics at the ILC}.
	\newblock 9 2007.
	
	\bibitem{Erler:2000jg}
	J.~Erler, S.~Heinemeyer, W.~Hollik, G.~Weiglein, and P.~M. Zerwas.
	\newblock {Physics impact of GigaZ}.
	\newblock {\em Phys. Lett. B}, 486:125--133, 2000.
	
	\bibitem{Sun:2013liv}
	Zhan Sun, Xing-Gang Wu, Gu~Chen, Jun Jiang, and Zhi Yang.
	\newblock {Heavy quarkonium production through the semi-exclusive $e^+e^-$
		annihilation channels round the $Z^0$ peak}.
	\newblock {\em Phys. Rev. D}, 87(11):114008, 2013.
	
	\bibitem{Frixione:1993yw}
	Stefano Frixione, Michelangelo~L. Mangano, Paolo Nason, and Giovanni Ridolfi.
	\newblock {Improving the Weizsacker-Williams approximation in electron - proton
		collisions}.
	\newblock {\em Phys. Lett.}, B319:339--345, 1993.
	
	\bibitem{Chu:2017mnm}
	Xiaoxuan Chu, Elke-Caroline Aschenauer, Jeong-Hun Lee, and Liang Zheng.
	\newblock {Photon structure studied at an Electron Ion Collider}.
	\newblock {\em Phys. Rev. D}, 96(7):074035, 2017.
	
	\bibitem{Binosi:2003yf}
	D.~Binosi and L.~Theussl.
	\newblock {JaxoDraw: A Graphical user interface for drawing Feynman diagrams}.
	\newblock {\em Comput. Phys. Commun.}, 161:76--86, 2004.
	
	\bibitem{Ma:1997bi}
	J.~P. Ma, B.~H.~J. McKellar, and C.~B. Paranavitane.
	\newblock {$J/\psi$ production at photon - photon colliders as a probe of the
		color octet mechanism}.
	\newblock {\em Phys. Rev. D}, 57:606--609, 1998.
	
	\bibitem{Japaridze:1998ss}
	George Japaridze and Avto Tkabladze.
	\newblock {Color octet contribution to J / psi production at a photon linear
		collider}.
	\newblock {\em Phys. Lett. B}, 433:139--146, 1998.
	
	\bibitem{Godbole:2001pj}
	R.~M. Godbole, D.~Indumathi, and M.~Kramer.
	\newblock {$J/\psi$ production through resolved photon processes at $e^{+}
		e^{-}$ colliders}.
	\newblock {\em Phys. Rev. D}, 65:074003, 2002.
	
	\bibitem{Qiao:2001wv}
	Cong-Feng Qiao.
	\newblock {Double J / psi production at photon colliders}.
	\newblock {\em Phys. Rev. D}, 64:077503, 2001.
	
	\bibitem{Kniehl:2002wd}
	Bernd~A. Kniehl, Caesar~P. Palisoc, and Lennart Zwirner.
	\newblock {Associated production of heavy quarkonia and electroweak bosons at
		present and future colliders}.
	\newblock {\em Phys. Rev. D}, 66:114002, 2002.
	
	\bibitem{Klasen:2003zn}
	M.~Klasen, B.~A. Kniehl, L.~N. Mihaila, and M.~Steinhauser.
	\newblock {Charmonium production in polarized high-energy collisions}.
	\newblock {\em Phys. Rev. D}, 68:034017, 2003.
	
	\bibitem{Artoisenet:2007qm}
	Pierre Artoisenet, Fabio Maltoni, and Tim Stelzer.
	\newblock {Automatic generation of quarkonium amplitudes in NRQCD}.
	\newblock {\em JHEP}, 02:102, 2008.
	
	\bibitem{Klasen:2001mi}
	M.~Klasen, Bernd~A. Kniehl, L.~Mihaila, and M.~Steinhauser.
	\newblock {$J/\psi$ plus dijet associated production in two photon collisions}.
	\newblock {\em Nucl. Phys. B}, 609:518--536, 2001.
	
	\bibitem{Klasen:2008mh}
	M.~Klasen and J.~P. Lansberg.
	\newblock {Perspectives for inclusive quarkonium production in photon-photon
		collisions at the LHC}.
	\newblock {\em Nucl. Phys. B Proc. Suppl.}, 179-180:226--231, 2008.
	
	\bibitem{Li:2009zzu}
	Rong Li and Kuang-Ta Chao.
	\newblock {Photoproduction of $J/psi$ in association with a $c \bar{c}$ pair}.
	\newblock {\em Phys. Rev. D}, 79:114020, 2009.
	
	\bibitem{Chen:2014xka}
	Gu~Chen, Xing-Gang Wu, Hai-Bing Fu, Hua-Yong Han, and Zhan Sun.
	\newblock {Photoproduction of heavy quarkonium at the ILC}.
	\newblock {\em Phys. Rev. D}, 90(3):034004, 2014.
	
	\bibitem{Sun:2015hhv}
	Zhan Sun, Xing-Gang Wu, and Hong-Fei Zhang.
	\newblock {Prompt $J/\psi$ production in association with a $c\bar{c}$ pair
		within the framework of nonrelativistic QCD via photon-photon collisions at
		the International Linear Collider}.
	\newblock {\em Phys. Rev. D}, 92(7):074021, 2015.
	
	\bibitem{Chen:2016hju}
	Zi-Qiang Chen, Long-Bin Chen, and Cong-Feng Qiao.
	\newblock {NLO QCD Corrections for $J/\psi+ c + \bar{c}$ Production in
		Photon-Photon Collision}.
	\newblock {\em Phys. Rev. D}, 95(3):036001, 2017.
	
	\bibitem{Yang:2020xkl}
	Hao Yang, Zi-Qiang Chen, and Cong-Feng Qiao.
	\newblock {NLO QCD corrections to exclusive quarkonium-pair production in
		photon\textendash{}photon collision}.
	\newblock {\em Eur. Phys. J. C}, 80(9):806, 2020.
	
	\bibitem{Yang:2022yxb}
	Hao Yang, Zi-Qiang Chen, and Cong-Feng Qiao.
	\newblock {NLO QCD corrections to pseudoscalar quarkonium production with two
		heavy flavors in photon-photon collision}.
	\newblock {\em Phys. Rev. D}, 105(9):094014, 2022.
	
	\bibitem{TodorovaNova:2001pt}
	S.~Todorova-Nova.
	\newblock {(Some of) recent gamma gamma measurements from LEP}.
	\newblock In {\em {Multiparticle dynamics. Proceedings, 31st International
			Symposium, ISMD 2001, Datong, China, September 1-7, 2001}}, pages 62--67,
	2001.
	
	\bibitem{Abdallah:2003du}
	J.~Abdallah et~al.
	\newblock {Study of inclusive J / psi production in two photon collisions at
		LEP-2 with the DELPHI detector}.
	\newblock {\em Phys. Lett.}, B565:76--86, 2003.
	
	\bibitem{Klasen:2001cu}
	M.~Klasen, Bernd~A. Kniehl, L.~N. Mihaila, and M.~Steinhauser.
	\newblock {Evidence for color octet mechanism from CERN LEP-2 $\gamma \gamma
		\to J/\psi$ + $X$ data}.
	\newblock {\em Phys. Rev. Lett.}, 89:032001, 2002.
	
	\bibitem{Butenschoen:2011yh}
	Mathias Butenschoen and Bernd~A. Kniehl.
	\newblock {World data of J/psi production consolidate NRQCD factorization at
		NLO}.
	\newblock {\em Phys. Rev.}, D84:051501, 2011.
	
	\bibitem{He:2019tig1}
	Zhi-Guo He and Bernd~A. Kniehl.
	\newblock {Perspectives of heavy-quarkonium production at FCC-ee}.
	\newblock In {\em {Theory report on the 11th FCC-ee workshop}}, pages 91--98,
	2019.
	
	\bibitem{Chen:2021tmf}
	An-Ping Chen, Yan-Qing Ma, and Hong Zhang.
	\newblock {A Short Theoretical Review of Charmonium Production}.
	\newblock {\em Adv. High Energy Phys.}, 2022:7475923, 2022.
	
	\bibitem{Ginzburg:1981vm}
	I.~F. Ginzburg, G.~L. Kotkin, V.~G. Serbo, and Valery~I. Telnov.
	\newblock {Colliding gamma e and gamma gamma Beams Based on the Single Pass
		Accelerators (of Vlepp Type)}.
	\newblock {\em Nucl. Instrum. Meth.}, 205:47--68, 1983.
	
	\bibitem{Telnov:1989sd}
	Valery~I. Telnov.
	\newblock {Problems of Obtaining $\gamma \gamma$ and $\gamma \epsilon$
		Colliding Beams at Linear Colliders}.
	\newblock {\em Nucl. Instrum. Meth. A}, 294:72--92, 1990.
	
	\bibitem{Gluck:1999ub}
	M.~Gluck, E.~Reya, and I.~Schienbein.
	\newblock {Radiatively generated parton distributions of real and virtual
		photons}.
	\newblock {\em Phys. Rev.}, D60:054019, 1999.
	\newblock [Erratum: Phys. Rev.D62,019902(2000)].
	
	\bibitem{Wang:2004du}
	Jian-Xiong Wang.
	\newblock {Progress in FDC project}.
	\newblock {\em Nucl. Instrum. Meth.}, A534:241--245, 2004.
	
	\bibitem{Bodwin:2002cfe}
	Geoffrey~T. Bodwin and Andrea Petrelli.
	\newblock {Order-$v^4$ corrections to $S$-wave quarkonium decay}.
	\newblock {\em Phys. Rev. D}, 66:094011, 2002.
	\newblock [Erratum: Phys.Rev.D 87, 039902 (2013)].
	
	\bibitem{Tanabashi:2018oca}
	M.~Tanabashi et~al.
	\newblock {Review of Particle Physics}.
	\newblock {\em Phys. Rev. D}, 98(3):030001, 2018.
	
	\bibitem{Wu:2019mky}
	Xing-Gang Wu, Jian-Ming Shen, Bo-Lun Du, Xu-Dong Huang, Sheng-Quan Wang, and
	Stanley~J. Brodsky.
	\newblock {The QCD renormalization group equation and the elimination of
		fixed-order scheme-and-scale ambiguities using the principle of maximum
		conformality}.
	\newblock {\em Prog. Part. Nucl. Phys.}, 108:103706, 2019.
	
	\bibitem{Eichten:1995ch}
	Estia~J. Eichten and Chris Quigg.
	\newblock {Quarkonium wave functions at the origin}.
	\newblock {\em Phys. Rev.}, D52:1726--1728, 1995.
	
\end{thebibliography}

\end{document}